# New Families of Solutions for the Space Time Fractional Burgers' Equation

Abaker A. Hassaballa [1,2], Ahmed M. A. Adam[1,3], Eltayeb A. Yousif [1,4], and Mohamed I. Nouh[5]

[1] Department of Mathematics, Faculty of Science, Northern Border University, Arar 91431, Saudi Arabia

[2] Department of Mathematics, College of Applied & Industrial Sciences, Bahri University, Khartoum, Sudan

[3] Faculty of Engineering, Alzaiem Alazhari University, Khartoum North 13311, Sudan

[4] Department of Applied Mathematics, Faculty of Mathematical Sciences, University of Khartoum, Khartoum 11111, Sudan

[5] Astronomy Department, National Research Institute of Astronomy and Geophysics (NRIAG), 11421 Helwan, Cairo, Egypt

**Abstract**

In this paper, the hyperbolic tangent function method is applied for constructing exact solutions for space-time conformal fractional Burger's equation. Furthermore, the space-time conformal fractional Burgers' equation is tested for the Painlevé property and consequently new numerous exact solutions are generated via Bäcklund transform.

## 1. Introduction

Recently, fractional Calculus has made a very important impact to the most fields of science, such as mathematics, engineering, physics, economics, etc. [1-21]. The applications of fractional calculus are contemporary [4, 19]. Historically, the fractional derivatives introduced in different ways, for example, Riemann–Liouville, Riesz, Caputo, Modified Riemann–Liouville [1, 19]. In the year (2014), a new fractional derivative is defined by the authors Khalil et al [21], named conformal fractional derivative (CFD). The definition and basic concepts of CFD is developed [22]. Furthermore, the interpretations of CFD in engineering and physics applications are introduced and discussed [23]. The definition of CFD of a function $f: (0, \infty) \to \mathbb{R}$ of order $\alpha$, where $0 < \alpha \leq 1$ is mainly given by the following limit

$$D^\alpha f(t) = \lim_{\varepsilon \to 0} \frac{f(t + \varepsilon t^{1-\alpha}) - f(t)}{\varepsilon} \tag{1.1}$$

The function $f$ is said to be $\alpha$- differentiable. The fractional derivative at $t = 0$ is given as

$$f^\alpha(0) = \lim_{t \to 0^+} D^\alpha f(t) \tag{1.2}$$

The advantages of CFD is that satisfying the properties of the classical integer derivatives [21-23]. Assume that $f$ and $g$ are $\alpha$- differentiable functions and $\lambda, a, b$ are constants, then the CFD satisfies the following properties:

i. $D^\alpha t^p = p t^{p-\alpha}, p \in \mathbb{R}$.

ii. $D^\alpha \lambda = 0$.

iii. $D^\alpha f(t) = t^{1-\alpha} \dfrac{df}{dt}$.

iv. $D^\alpha(af + bg) = aD^\alpha f + bD^\alpha g$.

v. $D^\alpha(fg) = fD^\alpha g + gD^\alpha f$.

vi. $D^\alpha \left(\dfrac{f}{g}\right) = \dfrac{gD^\alpha f - fD^\alpha g}{g^2}$.

vii. $D^\alpha \left(f(g(t))\right) = \dfrac{df}{dg} D^\alpha g(t) = t^{1-\alpha} \dfrac{df}{dg} \dfrac{dg}{dt}$.

In mathematical physics and many other phenomena in various fields of applied science are described by nonlinear models, particularly by nonlinear partial differential equations (NLPDEs), such as astrophysics, optics, fluid dynamics, mathematical biology, plasma physics and so on. It is important to search for the solutions of the concerned models to understand and interpret their physical mechanism and behavior. The search for exact solutions to NLPDE has become of great interest by mathematicians and physicists and they have exerted great efforts for that. There are many powerful and efficient methods for finding exact solutions have been introduced, such as the inverse scattering method [24, 25], Hirota method [26], Bäcklund transformation [27-29], Darboux transformation [30], hyperbolic tangent function method [31, 32], Jacobi elliptic function method [33, 34], truncated Painlevé expansion method [35-38], homogenous balance method [39, 40], and there are other various methods in the literature. The Bäcklund transformations (BT) are considered as powerful tools for integrable systems to relate NLPDEs and their solutions [27, 28, 29, 41, 42]. Up to now, the research is still devoted and ongoing for finding the BT, e.g from the Painlevé property [35-38], the Ablowitz-Kaup-Newell-Segur (AKNS) system [41], the nonclassical symmetries [42], etc.

One of the simple NLPDEs in mathematical physics is the Burger's equation that arising in many areas of science such as Navier-Stokes equations, traffic flow, and acoustics [43, 44]. Consider the space-time conformable fractional Burger's equation (CFBE)

$$D_t^\alpha u + u D_x^\alpha u = \sigma D_x^{\alpha\alpha} u, \tag{1.3}$$

where $\sigma$ is arbitrary constants, $\alpha \in (0,1]$, and $D_x^{\alpha\alpha} u = D_x^\alpha(D_x^\alpha u)$. In a few years ago, many researchers are introduced the solutions of CFBE by different methods. The Hopf-Cole transform is applied to a time CFBE, subsequently the approximate analytical solution is founded by applying a Homotopy Analysis Method [45]. Also, the exact solution is obtained by using Fourier transform [46]. The residual power series method is introduced for finding approximate solutions of a time CFBE [47]. Also, the residual power series method is used in finding the solution of the space-time conformable fractional KdV-Burgers equation [48]. The solution of regular and singular space-time coupled CFBEs is formulated by applying double Laplace transform [49].

The rest of the paper is organized as follows: In Sec. 2, we show that the space-time CFBE possesses the Painlevé property. In Sec. 3 the exact solutions for the space-time CFBE is constructed based on the hyperbolic tangent function method. In Sec. 4, the BT is used for generating abundant new exact solutions for the space-time CFBE. The paper is concluded in Sec. 5.

## 2. Painlevé property for the space-time conformal fractional Burger's equation

In this section, we intent to test the Painlevé property for the space-time CFBE given by Eq. (1.3) following the approach introduced in the Refs. [35- 37, 50]. The space-time CFBE has the Painlevé property when all the movable singularities are simple poles. For Eq. (1.3) we let

$$u = u\left(\frac{x^\alpha}{\alpha}, \frac{t^\alpha}{\alpha}\right) = \phi^n \sum_{j=0}^{\infty} \phi^j u_j, \tag{2.1}$$

where $u_0 \neq 0$, $n$ is an integer, $\phi = \phi\left(\frac{x^\alpha}{\alpha}, \frac{t^\alpha}{\alpha}\right)$ and $u_j = u_j\left(\frac{x^\alpha}{\alpha}, \frac{t^\alpha}{\alpha}\right)$ are analytic functions of $\left(\frac{x^\alpha}{\alpha}, \frac{t^\alpha}{\alpha}\right)$ in a neighborhood of $M = \left\{\left(\frac{x^\alpha}{\alpha}, \frac{t^\alpha}{\alpha}\right) : \phi\left(\frac{x^\alpha}{\alpha}, \frac{t^\alpha}{\alpha}\right) = 0\right\}$.

To determine the values of $n$ we consider the ansätz

$$u = \phi^n u_0 \tag{2.2}$$

By using chain rule on Eq. (2.2) we obtain

$$D_t^\alpha u = n u_0 \phi^{n-1} D_t^\alpha \phi + \phi^n D_t^\alpha u_0$$
$$D_x^\alpha u = n u_0 \phi^{n-1} D_x^\alpha \phi + \phi^n D_x^\alpha u_0 \qquad (2.3)$$
$$D_x^{\alpha\alpha} u = n u_0 \phi^{n-1} D_x^{\alpha\alpha} \phi + n(n-1) u_0 \phi^{n-2} (D_x^\alpha \phi)^2 + 2n \phi^{n-1} D_x^\alpha \phi D_x^\alpha u_0 + \phi^n D_x^{\alpha\alpha} u_0.$$

By substituting Eqs. (2.2) and (2.3) into Eq. (1.3), we get

$$n u_0 \phi^{n-1} D_t^\alpha \phi + \phi^n D_t^\alpha u_0 + n u_0^2 \phi^{2n-1} D_x^\alpha \phi + u_0 \phi^{2n} D_x^\alpha u_0$$
$$= n\sigma u_0 \phi^{n-1} D_x^{\alpha\alpha} \phi + n(n-1) \sigma u_0 \phi^{n-2} (D_x^\alpha \phi)^2 + 2n\sigma \phi^{n-1} D_x^\alpha \phi D_x^\alpha u_0 + \sigma \phi^n D_x^{\alpha\alpha} u_0. \qquad (2.4)$$

The dominant terms of Eq. (2.4) are $\phi^{2n-1}$ & $\phi^{n-2}$. Balancing of these gives $n = -1$. Thus,

$$u_0 = -2\sigma D_x^\alpha \phi \qquad (2.5)$$

Back to Eq. (2.1), we can consider the following Ansätz with resonance $r$ as

$$u = u_0 \phi^{-1} + b\phi^{r-1} = -2\sigma \phi^{-1} D_x^\alpha \phi + b\phi^{r-1} \qquad (2.6)$$

From Eq. (2.6) we get

$$D_x^\alpha u = 2\sigma \phi^{-2} (D_x^\alpha \phi)^2 - 2\sigma \phi^{-1} D_x^{\alpha\alpha} \phi + b(r-1) \phi^{r-2} D_x^\alpha \phi$$
$$= 2\sigma \phi^{-2} (D_x^\alpha \phi)^2 + b(r-1) \phi^{r-2} D_x^\alpha \phi + \cdots$$
$$D_x^{\alpha\alpha} u = -4\sigma \phi^{-3} (D_x^\alpha \phi)^3 + 4\sigma \phi^{-2} D_x^\alpha \phi D_x^{\alpha\alpha} \phi + 2\sigma \phi^{-2} D_x^\alpha \phi D_x^{\alpha\alpha} \phi - 2\sigma \phi^{-1} D_x^{\alpha\alpha\alpha} \phi \qquad (2.7)$$
$$\quad + b(r-1)(r-2) \phi^{r-3} (D_x^\alpha \phi)^2 + b(r-1) \phi^{r-2} D_x^{\alpha\alpha} \phi$$
$$= -4\sigma \phi^{-3} (D_x^\alpha \phi)^3 + b(r-1)(r-2) \phi^{r-3} (D_x^\alpha \phi)^2 + \cdots$$

The fractional derivatives are rearranged in terms of powers of $D_x^\alpha \phi$. Now. By substituting Eqs. (2.6) and (2.7) into $u D_x^\alpha u = \sigma D_x^{\alpha\alpha} u$, we have

$$(-2\sigma \phi^{-1} D_x^\alpha \phi + b\phi^{r-1})(2\sigma \phi^{-2} (D_x^\alpha \phi)^2 + b(r-1)\phi^{r-2} D_x^\alpha \phi + \cdots)$$
$$= \sigma(-4\sigma \phi^{-3}(D_x^\alpha \phi)^3 + b(r-1)(r-2)\phi^{r-3}(D_x^\alpha \phi)^2 + \cdots)$$
$$\Rightarrow -2b(r-1)\sigma \phi^{r-3}(D_x^\alpha \phi)^2 + 2b\sigma \phi^{r-3}(D_x^\alpha \phi)^2 = \sigma b(r-1)(r-2)\phi^{r-3}(D_x^\alpha \phi)^2 + \cdots$$
$$\Rightarrow (-2r + 2 + 2 - r^2 + 3r - 2)\sigma b \phi^{r-3}(D_x^\alpha \phi)^2 = \cdots$$
$$\Rightarrow -(r+1)(r-2)\sigma b \phi^{r-3}(D_x^\alpha \phi)^2 = \cdots$$
$$\Rightarrow r = -1 \text{ and } r = 2.$$

If $M$ is a singularity manifold, it is obtained that $n = -1$. By leading order analysis,

$$u = \phi^{-1} \sum_{j=0}^{\infty} \phi^j u_j = \sum_{j=0}^{\infty} \phi^{j-1} u_j, \qquad (2.8)$$

From (2.8) we get

$$D_t^\alpha u = \left[ \sum_{j=0}^{\infty} [(j-1)\phi^{j-2} u_j D_t^\alpha \phi + \phi^{j-1} D_t^\alpha u_j] \right] = \left[ \sum_{j=0}^{\infty} [(j-2)\phi^{j-3} u_{j-1} D_t^\alpha \phi + \phi^{j-3} D_t^\alpha u_{j-2}] \right],$$

$$D_x^\alpha u = \left[\sum_{j=0}^{\infty}[(j-1)\phi^{j-2}u_j D_x^\alpha \phi + \phi^{j-1}D_x^\alpha u_j]\right] = \left[\sum_{j=0}^{\infty}[(j-1)\phi^{j-2}u_j D_x^\alpha \phi + \phi^{j-2}D_x^\alpha u_{j-1}]\right],$$

$$D_x^{\alpha\alpha} u = \left[\sum_{j=0}^{\infty}[(j-1)(j-2)u_j \phi^{j-3}(D_x^\alpha \phi)^2 + (j-1)\phi^{j-2}u_j D_x^{\alpha\alpha}\phi + 2(j-1)\phi^{j-2}D_x^\alpha \phi D_x^\alpha u_j \right.$$

$$\left. + \phi^{j-1}D_x^{\alpha\alpha} u_j]\right],$$

$$= \left[\sum_{j=0}^{\infty}[(j-1)(j-2)u_j \phi^{j-3}(D_x^\alpha \phi)^2 + (j-2)\phi^{j-3}u_{j-1}D_x^{\alpha\alpha}\phi \right.$$

$$\left. + 2(j-2)\phi^{j-3}D_x^\alpha \phi D_x^\alpha u_{j-1} + \phi^{j-3}D_x^{\alpha\alpha} u_{j-2}]\right].$$

The recursion relations for $u_j$ are found to be

$$D_t^\alpha u_{j-2} + (j-2)u_{j-1}D_t^\alpha \phi + \sum_{m=0}^{j} u_{j-m}[D_x^\alpha u_{m-1} + (m-1)u_m D_x^\alpha \phi]$$

$$= \sigma[(j-1)(j-2)u_j (D_x^\alpha \phi)^2 + (j-2)u_{j-1}D_x^{\alpha\alpha}\phi + 2(j-2)D_x^\alpha \phi D_x^\alpha u_{j-1} + D_x^{\alpha\alpha} u_{j-2}], (2.9)$$

where $u_k = 0$ for $k = -1, -2, -3, \ldots$

From the recurrence formula (2.9) we obtain for:

$j = 0 \Rightarrow -u_0^2 D_x^\alpha \phi = 2\sigma u_0 (D_x^\alpha \phi)^2$

$\Rightarrow u_0 = -2\sigma D_x^\alpha \phi.$  (2.10)

Eq. (2.10) is identical to Eq. (2.5).

$$j = 1 \Rightarrow -u_0 D_t^\alpha \phi + \sum_{m=0}^{1} u_{1-m}[D_x^\alpha u_{m-1} + (m-1)u_m D_x^\alpha \phi] = -\sigma[u_0 D_x^{\alpha\alpha}\phi + 2D_x^\alpha \phi D_x^\alpha u_0]$$

$\Rightarrow -u_0 D_t^\alpha \phi - u_0 u_1 D_x^\alpha \phi + u_0 D_x^\alpha u_0 = -\sigma[u_0 D_x^{\alpha\alpha}\phi + 2D_x^\alpha \phi D_x^\alpha u_0]$

$$= \sigma[2\sigma D_x^\alpha \phi D_x^{\alpha\alpha}\phi + 4\sigma D_x^\alpha \phi D_x^{\alpha\alpha}\phi] = 6\sigma^2 D_x^\alpha \phi D_x^{\alpha\alpha}\phi$$

$\Rightarrow -u_0 D_t^\alpha \phi - u_0 u_1 D_x^\alpha \phi + 4\sigma^2 D_x^\alpha \phi D_x^{\alpha\alpha}\phi = 6\sigma^2 D_x^\alpha \phi D_x^{\alpha\alpha}\phi$

$\Rightarrow -u_0 D_t^\alpha \phi - u_0 u_1 D_x^\alpha \phi = 2\sigma^2 D_x^\alpha \phi D_x^{\alpha\alpha}\phi = -u_0 \sigma D_x^{\alpha\alpha}\phi \Rightarrow$

$\Rightarrow D_t^\alpha \phi + u_1 D_x^\alpha \phi - \sigma D_x^{\alpha\alpha}\phi = 0$  (2.11)

$j = 2$ and considering Eqs. (2.10, 2.11), then from Eq. (2.9) we get

$$D_t^\alpha u_0 + \sum_{m=0}^{2} u_{2-m}[D_x^\alpha u_{m-1} + (m-1)u_m D_x^\alpha \phi] = \sigma D_x^{\alpha\alpha} u_0$$

$$\Rightarrow D_t^\alpha u_0 + u_1 D_x^\alpha u_0 + u_0 D_x^\alpha u_1 = \sigma D_x^{\alpha\alpha} u_0$$

$$\Rightarrow D_x^\alpha (D_t^\alpha \phi + u_1 D_x^\alpha \phi - \sigma D_x^{\alpha\alpha} \phi) = 0 \tag{2.12}$$

By Eq. (2.11) the compatability condition Eq. (2.12) at $j = 2$ is satisfied identically. For $j = 3$, then from recurrence formula given by Eq. (2.9) we have

$$D_t^\alpha u_1 + u_2 D_t^\alpha \phi + \sum_{m=0}^{3} u_{3-m}[D_x^\alpha u_{m-1} + (m-1)u_m D_x^\alpha \phi]$$
$$= \sigma[2u_3(D_x^\alpha \phi)^2 + u_2 D_x^{\alpha\alpha} \phi + 2D_x^\alpha \phi D_x^\alpha u_2 + D_x^{\alpha\alpha} u_1]$$

$$\Rightarrow [D_t^\alpha u_1 + u_1 D_x^\alpha u_1 - \sigma D_x^{\alpha\alpha} u_1] + u_2[D_t^\alpha \phi + u_1 D_x^\alpha \phi - \sigma D_x^{\alpha\alpha} \phi]$$
$$+ [u_2 D_x^\alpha u_0 + u_0 D_x^\alpha u_2 - 2\sigma D_x^\alpha \phi D_x^\alpha u_2] + u_3 D_x^\alpha \phi[u_0 - 2\sigma D_x^\alpha \phi] = 0$$

$$\Rightarrow [D_t^\alpha u_1 + u_1 D_x^\alpha u_1 - \sigma D_x^{\alpha\alpha} u_1] + [u_2 D_x^\alpha u_0 + 2u_0 D_x^\alpha u_2] + 2u_3 u_0 D_x^\alpha \phi = 0 \tag{2.13}$$

Since the resonances occur at $r = -1, 2$, and $(\phi, u_2)$ are arbitrary functions of $\left(\frac{x^\alpha}{\alpha}, \frac{t^\alpha}{\alpha}\right)$ in the expansion (2.13). If we let the arbitrary functions $u_2 = u_3 = 0$, then we get

$$D_t^\alpha u_1 + u_1 D_x^\alpha u_1 = \sigma D_x^{\alpha\alpha} u_1 \tag{2.14}$$

For $j = 3$ then Eq. (2.14) is automatically satisfied for $u_1$.

for $j = 4$, $u_0 = -2\sigma D_x^\alpha \phi$, and $u_2 = u_3 = 0$ then from Eq. (2.9) we obtain

$$\sum_{m=0}^{4} u_{4-m}[D_x^\alpha u_{m-1} + (m-1)u_m D_x^\alpha \phi] = 6\sigma u_4 (D_x^\alpha \phi)^2$$

$$\Rightarrow u_4 D_x^\alpha \phi (2u_0 - 6\sigma D_x^\alpha \phi) = 0 \Rightarrow u_4 = 0.$$

Thus, we conclude that all

$$u_j = 0, j \geq 2, \tag{2.15}$$

providing $u_1$ satisfies space-time CFBE (Eq. (2.14)). Hence, the space-time CFBE possesses the Painlevé property and the truncation of the Painlevé expansion Eq. (2.8) then takes the form

$$u = -\frac{2\sigma}{\phi} D_x^\alpha \phi + u_1, \tag{2.16}$$

which is the Bäcklund transform for the space-time CFBE. When $u_1 = 0$ then from Eq. (2.16) we obtain

$$u = -\frac{2\sigma}{\phi} D_x^\alpha \phi, \tag{2.17}$$

Eq. (2.17) yields the fractional Cole-Hopf transform. When $u_1 = \phi$ then we have

$$u = -\frac{2\sigma}{\phi} D_x^\alpha \phi + \phi, \tag{2.18}$$

where

$$D_t^\alpha \phi + \phi D_x^\alpha \phi = \sigma D_x^{\alpha\alpha} \phi \tag{2.19}$$

Eqs. (2.18) and (2.19) represent the Bäcklund transform for the space-time CFBE.

## 3. Exact solutions for the space-time conformal fractional Burger's equation

The hyperbolic tangent function method (tanh-method) is an efficient method that used for constructing the exact traveling wave solutions for NLPDEs. The Ansätz is considered as a power series in tanh, where tanh is introduced as a new variable. Moreover, the derivatives of tanh are also given in terms of tanh itself [31, 32]. In the following steps we seek to describe the general tanh-method for constructing the solutions of the space-time conformal fractional partial deferential equations (CFPDEs). Consider a general space-time conformal fractional partial differential equation (space-time CFPDE):

$$H(u, D_t^\alpha u, D_x^\alpha u, D_t^{\alpha\alpha} u, D_t^\alpha(D_x^\alpha u), D_x^{\alpha\alpha} u, \dots) = 0. \tag{3.1}$$

In the following steps, we summarize the tanh-method for solving Eq. (3.1):

Step 1: consider the traveling wave solution of Eq. (3.1) as

$$u(x,t) = u(\xi) \qquad \xi = \frac{k}{\alpha}(x^\alpha - \omega t^\alpha), \tag{3.2}$$

where $k$ and $\omega$ are the wave number and wave velocity, respectively. Substitution of Eq. (3.2) into Eq. (3.1) produces the following ordinary differential equation for $u(\xi)$

$$\widetilde{H}(u, u', u'', \dots) = 0 \qquad u' = \frac{du}{d\xi}, \dots etc. \tag{3.3}$$

Step 2: Assume that the solution of Eq. (3.3) can be expressed as a finite power series of $F(\xi)$

$$u(\xi) = a_0 + \sum_{j=1}^{s} a_j F^j(\xi), \qquad a_s \neq 0, \tag{3.4}$$

where $s \in \mathbb{N}$, which is determined by balancing the highest power of the linear term with the highest power of the nonlinear term in Eq. (3.3), and $a_j$ are constants to be determined. The new exact solutions of the space-time CFPDE can be obtained via the solutions of the Riccati equation that satisfied by tanh function. Consider the required Riccati equation to be

$$F' = A + BF + CF^2, \quad ' \equiv \frac{d}{d\xi}, \tag{3.5}$$

where $A, B$ and $C$ are constants.

Step 3: Substitution of Eq. (3.4) into Eq. (3.3), generates system of algebraic equations for $a_0, a_1, \ldots, a_s, \omega,$ and $k$.

Step 4: Solution of the system obtained in step 3, produces the values of $a_0, a_1, \ldots, a_s, \omega$ and $k$ in terms of $A, B$ and $C$. By substituting these results into Eq. (3.4), we obtain the general form of traveling wave solution of Eq. (3.1).

Choosing of each proper value for $A, B$ and $C$ in Eq. (3.5) corresponds a solution $F(\xi)$ of Eq. (3.5) that is could be one of the hyperbolic function or triangular function as follows.

Case 1: If $A = C = 1$, and $B = 0$, then Eq. (3.5) has the solutions, $tan\xi$.

Case 2: If $A = C = -1$, and $B = 0$, then Eq. (3.5) has the solutions, $cot\xi$.

Case 3: If $A = 1, B = 0$, and $C = -1$, then Eq. (3.5) has the solutions, $tanh\xi, coth\xi$.

Case 4: If $A = \frac{1}{2}, B = 0$ and, $C = -\frac{1}{2}$, then Eq. (3.5) has the solutions, $tanh\xi \pm i\, sech\xi$, $coth\xi \pm csch\xi, \frac{tanh\xi}{1 \pm sech\xi}, \frac{coth\xi}{1 \pm i\, csch\xi}, i^2 = -1.$

Case 5: If $A = C = \frac{1}{2}$, and $B = 0$, then Eq. (3.5) has the solutions, $tan\xi \pm sec\xi,\ csc\xi - cot\xi$, $\frac{tan\xi}{1 \pm sec\xi}$.

Case 6: If $A = C = -\frac{1}{2}$, and $B = 0$, then Eq. (3.5) has the solutions, $cot\xi \pm csc\xi,\ sec\xi - tan\xi$, $\frac{cot\xi}{1 \pm csc\xi}$.

Case 7: If $A = 1, C = -4$, and $B = 0$, then Eq. (3.5) has the solutions, $\frac{tanh\xi}{1+tanh^2\xi}$.

Case 8: If $A = 1, C = 4$, and $B = 0$, then Eq. (3.5) has the solutions, $\frac{tan\xi}{1-tan^2\xi}$.

Case 9: If $A = -1, C = -4$, and $B = 0$, then Eq. (3.5) has the solutions, $\frac{cot\xi}{1-cot^2\xi}$.

Case 10: If $A = 1, B = -2$, and $C = 2$, then Eq. (3.5) has the solutions, $\frac{tan\xi}{1+tan\xi}$.

Case 11: If $A = 1$ and $B = C = 2$, then Eq. (3.5) has the solutions, $\frac{tan\xi}{1-tan\xi}$.

Case 12: If $A = -1, B = 2$, and $C = -2$, then Eq. (3.5) has the solutions, $\frac{cot\xi}{1+cot\xi}$.

Case 13: If $A = -1, B = C = -2$, then Eq. (3.5) has the solutions, $\frac{cot\xi}{1-cot\xi}$.

Case 14: If $A = B = 0$ and $C \neq 0$, then Eq. (3.5) has the solutions, $\frac{-1}{C\xi + C_0}$.

Case 15: If $A \neq 0$, $C = 0$, and $B \neq 0$, then Eq. (3.5) has the solutions, $\frac{1}{B}(\exp(B\xi) - A)$.

Case 16: If $A = 0$ and $B = C = 1$, then Eq. (3.5) has the solutions, $\frac{\exp(\xi)}{1-\exp(\xi)}$.

Case 17: If $A = 0$, and $B = C = \frac{1}{2}$, then Eq. (3.5) has the solutions, $\frac{\exp\left(\frac{\xi}{2}\right)}{1-\exp\left(\frac{\xi}{2}\right)}$.

Case 18: If $A = -\frac{1}{2}$, $B = 0$, and $C = \frac{1}{2}$, then Eq. (3.5) has the solutions, $-\tanh\left(\frac{\xi}{2}\right), -\coth\left(\frac{\xi}{2}\right)$.

Case 19: If $A = -1$, and $B = C = 2$, then Eq. (3.5) has the solutions, $-\frac{1}{2} - \frac{\sqrt{3}\tanh(\sqrt{3}\xi)}{2}$,

$-\frac{1}{2} - \frac{\sqrt{3}\coth(\sqrt{3}\xi)}{2}$.

Case 20: If $A = 1$, $B = 1$, and $C = 1$, then Eq. (3.5) has the solutions, $-\frac{1}{2} + \frac{\sqrt{3}\tan\left(\frac{\sqrt{3}}{2}\xi\right)}{2}$,

$-\frac{1}{2} - \frac{\sqrt{3}\cot\left(\frac{\sqrt{3}}{2}\xi\right)}{2}$.

Case 21: If $A = -4$, $B = 0$, and $C = 4$, then Eq. (3.5) has the solutions, $-\tanh(4\xi), -\coth(4\xi)$.

Case 22: If $A = \frac{1}{2}$, $B = -1$, and $C = 1$, then Eq. (3.5) has the solutions, $\frac{1}{2} + \frac{1}{2}\tan\left(\frac{\xi}{2}\right)$,

$\frac{1}{2} - \frac{1}{2}\cot\left(\frac{\xi}{2}\right)$.

Now, we need to implement the tanh-method technique into the space-time CFBE given by Eq. (1.3) to generate new exact solutions. Firstly, substituting the traveling wave solution given by Eq. (3.2) into Eq. (1.3) we obtain

$$-\omega u' + uu' - k\sigma u'' = 0, \qquad (3.6)$$

where $D_t^\alpha u = -\omega k u'$, $D_x^\alpha u = ku'$, $D_x^{\alpha\alpha} u = k^2 u''$, by balancing $u''$ with $uu'$ gives $s = 1$. Use $s = 1$ in Eq. (3.4), then the solution of Eq. (1.3) can be expressed as

$$u = a_0 + a_1 F, \qquad (3.7)$$

substituting Eq. (3.7) into Eq. (3.6) and using Eq. (3.5), then we obtain a set of algebraic equations with respect to $F^i (i = 0,1,2,3)$. Equating the coefficients of $F^i (i = 0,1,2,3)$ to zero. The solution of the resulting system is given by

$$a_0 = \omega + \sigma B k, \quad a_1 = 2\sigma C k, \qquad (3.8)$$

with $\omega$ and $k$ are arbitrary constants. Inserting Eq. (3.8) into Eq. (3.7) and using the special solutions of Eq. (3.5), we obtain the following soliton like-solution and triangular periodic solutions of the space-time CFBE:

$$u_1 = \omega + 2\sigma k \tan\xi, \tag{3.9}$$

$$u_2 = \omega - 2\sigma k \cot\xi, \tag{3.10}$$

$$u_3 = \omega - 2\sigma k \tanh\xi, \tag{3.11}$$

$$u_4 = \omega - 2\sigma k \coth\xi, \tag{3.12}$$

$$u_5 = \omega - \sigma k(\tanh\xi \pm i\,\text{sech}\xi), \tag{3.13}$$

$$u_6 = \omega - \sigma k(\coth\xi \pm \text{csch}\xi), \tag{3.14}$$

$$u_7 = \omega + \sigma k(\tan\xi \pm \sec\xi), \tag{3.15}$$

$$u_8 = \omega - \sigma k(\cot\xi \pm \csc\xi), \tag{3.16}$$

$$u_9 = \omega - \frac{8\sigma k \tanh\xi}{1+\tanh^2\xi}, \tag{3.17}$$

$$u_{10} = \omega + \frac{8\sigma k \tan\xi}{1-\tan^2\xi}, \tag{3.18}$$

$$u_{11} = \omega - \frac{8\sigma k \cot\xi}{1-\cot^2\xi}, \tag{3.19}$$

$$u_{12} = \omega - \frac{\sigma k \tanh\xi}{1+\text{sech}\xi}, \tag{3.20}$$

$$u_{13} = \omega - \frac{\sigma k \coth\xi}{1+i\,\text{csch}\xi}, \tag{3.21}$$

$$u_{14} = \omega + \frac{\sigma k \tan\xi}{1+\sec\xi}, \tag{3.22}$$

$$u_{15} = \omega - \frac{\sigma k \cot\xi}{1+\csc\xi}, \tag{3.23}$$

$$u_{16} = \omega - 2\sigma k + \frac{4\sigma k \tan\xi}{1+\tan\xi}, \tag{3.24}$$

$$u_{17} = \omega + 2\sigma k - \frac{4\sigma k \cot\xi}{1+\cot\xi}, \tag{3.25}$$

$$u_{18} = \frac{(\sigma k+\omega)+(\sigma k-\omega)e^\xi}{1-e^\xi}, \tag{3.26}$$

$$u_{19} = \frac{(\sigma k/2+\omega)+(\sigma k/2-\omega)e^{\xi/2}}{1-e^{\xi/2}}, \tag{3.27}$$

$$u_{20} = \omega - \sigma k \tanh\frac{\xi}{2}, \tag{3.28}$$

$$u_{21} = \omega - \sigma k \coth\frac{\xi}{2}, \tag{3.29}$$

$$u_{22} = \omega - 2\sqrt{3}\sigma k \tanh\sqrt{3}\xi, \tag{3.30}$$

$$u_{23} = \omega - 2\sqrt{3}\sigma k\, cothh\sqrt{3}\xi, \tag{3.31}$$

$$u_{24} = \omega + \sqrt{3}\sigma k\, tan\left(\tfrac{\sqrt{3}}{2}\xi\right), \tag{3.32}$$

$$u_{25} = \omega - \sqrt{3}\sigma k\, cot\left(\tfrac{\sqrt{3}}{2}\xi\right), \tag{3.33}$$

$$u_{26} = \omega - 8\sigma k\, tanh(4\xi), \tag{3.34}$$

$$u_{27} = \omega - 8\sigma k\, coth(4\xi), \tag{3.35}$$

$$u_{28} = \omega + \sigma k\, tan\left(\tfrac{\xi}{2}\right), \tag{3.36}$$

$$u_{29} = \omega - \sigma k\, cot\left(\tfrac{\xi}{2}\right), \tag{3.37}$$

Remarks:

1- Making the transformation $k \to 2k$ then Eq. (3.28) and Eq. (3.29) transformed to Eq. (3.11) and Eq. (3.12) respectively, and if $k \to 2ki$ where $i = \sqrt{-1}$, they transformed to Eq. (3.9) and Eq. (3.10) respectively.

2- Making the transformation $k \to \frac{k}{\sqrt{3}}$ then Eq. (3.30) and Eq. (3.31) transformed to Eq. (3.11) and Eq. (3.12) respectively.

3- Making the transformation $k \to \frac{2k}{\sqrt{3}}$ then Eq. (3.32) and Eq. (3.33) transformed to Eq. (3.9) and Eq. (3.10) respectively.

4- Making the transformation $k \to 2k$ then Eq. (3.27) transformed to Eq. (3.26).

5- Making the transformation $k \to \frac{k}{4}$ then Eq. (3.34) and Eq. (3.35) transformed to Eq. (3.11) and Eq. (3.12) respectively.

6- Making the transformation $k \to 2k$ then Eq. (3.36) and Eq. (3.37) transformed to Eq. (3.9) and Eq. (3.10) respectively.

Also, we can get two new exact solutions

$$u_{30} = \omega - 2\sigma k\, (tanh\xi + coth\xi), \tag{3.38}$$

$$u_{31} = \omega + 2\sigma k\, (tan\xi - cot\xi), \tag{3.39}$$

with $\xi = \frac{k}{\alpha}(x^\alpha - \omega t^\alpha)$.

In the following section we use the obtained solutions Eqs. (3.9) - (3.39) to generate new abundant exact solutions via BT.

## 4. BT and abundant exact solutions

From section 2, we show that the space-time CFBE possesses the Painlevé property and it has a BT in the form

$$u = -\frac{2\sigma}{\phi} D_x^\alpha \phi + w, \tag{4.1}$$

where $\phi = \phi\left(\frac{x^\alpha}{\alpha}, \frac{t^\alpha}{\alpha}\right)$ is the singular manifold variable, $w$ is a function of $\frac{x^\alpha}{\alpha}$ and $\frac{t^\alpha}{\alpha}$. Also, the function $w$ solves the space-time CFBE given by Eq. (1.3) and the function $\phi$ satisfies the FDE

$$D_t^\alpha \phi + w D_x^\alpha \phi = \sigma D_x^{\alpha\alpha} \phi. \tag{4.2}$$

Now, if we take $w = \phi$ then the function $\phi$ satisfies also the space-time CFBE

$$D_t^\alpha \phi + \phi D_x^\alpha \phi = \sigma D_x^{\alpha\alpha} \phi, \tag{4.3}$$

thus the BT for the space-time CFBE takes the following recurrence form

$$u_{n+1} = -\frac{2\sigma}{u_n} \frac{\partial^\alpha u_n}{\partial x^\alpha} + u_n. \tag{4.4}$$

We turn to the application of the BT for the FDEs. Their power lies in that they may be used to generate additional solutions of the FDEs. Here $u_{n+1}$ quantities refer to new solution and $u_n$ quantities refer to old solution. This means that, on the basis of a known solution to the space-time CFBE, we are able to find new solution of space-time CFBE. To construct the new solution of the space-time CFBE one can start with the solution $u_1$ obtained in Eq. (3.9) and using BT given in Eq. (4.4) we get the following set of new solutions:

$$u_{32} = \frac{-4\sigma^2 k^2 + \omega^2 + 4\sigma k\omega \tan\xi}{\omega + 2\sigma k \tan\xi}, \tag{4.5}$$

Inserting $u_{32}$ into the BT given in Eq. (4.4) we have

$$u_{33} = \frac{12\sigma^2 k^2 \omega - \omega^3 + 2\sigma k(4\sigma^2 k^2 - 3\omega^2) \tan\xi}{4\sigma^2 k^2 - \omega^2 - 4\sigma k\omega \tan\xi}, \tag{4.6}$$

furthermore, using $u_{33}$ and BT given in Eq. (4.4) we get

$$u_{34} = \frac{-16\sigma^4 k^4 + 24\sigma^2 k^2 \omega^2 - \omega^4 + (32\sigma^3 k^3 \omega - 8\omega^3 \sigma k) \tan\xi}{12\sigma^2 k^2 \omega - \omega^3 + 2\sigma k(4\sigma^2 k^2 - 3\omega^2) \tan\xi}, \tag{4.7}$$

and so on, we can get a sequences of exact solutions generated by the known tan-function solution Eq. (3.9) of the space-time CFBE. Starting from $u_2$ obtained in Eq. (3.10) and using BT given in Eq. (4.4) we get

$$u_{35} = \frac{-4\sigma^2 k^2 + \omega^2 - 4\sigma k\omega \cot\xi}{\omega - 2\sigma k \cot\xi}, \tag{4.8}$$

Inserting $u_{35}$ into the BT given in Eq. (4.4) we have

$$u_{36} = \frac{-12\sigma^2 k^2\omega+\omega^3+2\sigma k(4\sigma^2 k^2-3\omega^2)\cot\xi}{-4\sigma^2 k^2+\omega^2-4\sigma k\omega \cot\xi}, \tag{4.9}$$

furthermore, using $u_{36}$ and BT given in Eq. (4.4) we get

$$u_{37} = \frac{16\sigma^4 k^4-24\sigma^2 k^2\omega^2+\omega^4+(32\sigma^3 k^3\omega-8\omega^3\sigma k)\cot\xi}{-12\sigma^2 k^2\omega+\omega^3+2\sigma k(4\sigma^2 k^2-3\omega^2)\cot\xi}, \tag{4.10}$$

and so on, we can get a sequences of exact solutions generated by the cot-function solution Eq. (3.10) of the space time fractional CFBE. Starting from $u_3$ obtained in Eq. (3.11) and using BT given in Eq. (4.4) we get

$$u_{38} = \frac{-4\sigma^2 k^2-\omega^2+4\sigma k\omega\tanh\xi}{-\omega+2\sigma k\tanh\xi}, \tag{4.11}$$

Inserting $u_{38}$ into the BT given in Eq. (4.4) we have

$$u_{39} = \frac{12\sigma^2 k^2\omega+\omega^3-2\sigma k(4\sigma^2 k^2+3\omega^2)\tanh\xi}{4\sigma^2 k^2+\omega^2-4\sigma k\omega\tanh\xi}, \tag{4.12}$$

furthermore, using $u_{39}$ and BT given in Eq. (4.4) we get

$$u_{40} = \frac{16\sigma^4 k^4+24\sigma^2 k^2\omega^2+\omega^4-(32\sigma^3 k^3\omega+8\omega^3\sigma k)\tanh\xi}{12\sigma^2 k^2\omega+\omega^3-2\sigma k(4\sigma^2 k^2+3\omega^2)\tanh\xi}, \tag{4.13}$$

and so on, we can get a sequences of exact solutions generated by the tanh-function solution Eq. (3.11) of the space-time CFBE. Starting from $u_4$ obtained in Eq. (3.12) and using BT given in Eq. (4.4) we get

$$u_{41} = \frac{-4\sigma^2 k^2-\omega^2+4\sigma k\omega\coth\xi}{-\omega+2\sigma k\coth\xi}, \tag{4.14}$$

Inserting $u_{41}$ into the BT given in Eq. (4.4) we have

$$u_{42} = \frac{12\sigma^2 k^2\omega+\omega^3-2\sigma k(4\sigma^2 k^2+3\omega^2)\coth\xi}{4\sigma^2 k^2+\omega^2-4\sigma k\omega\coth\xi}, \tag{4.15}$$

furthermore, using $u_{42}$ and BT given in Eq. (4.4) we get

$$u_{43} = \frac{16\sigma^4 k^4+24\sigma^2 k^2\omega^2+\omega^4-(32\sigma^3 k^3\omega+8\omega^3\sigma k)\coth\xi}{12\sigma^2 k^2\omega+\omega^3-2\sigma k(4\sigma^2 k^2+3\omega^2)\coth\xi}, \tag{4.16}$$

and so on, we can get a sequences of exact solutions generated by the coth-function solution Eq. (3.12) of the space-time CFBE. Starting from $u_5$ obtained in Eq. (3.13) and using BT given in Eq. (4.4) we get

$$u_{44} = \frac{\sigma^2 k^2+\omega^2-2\sigma k\omega\,(\tanh\xi\pm i\,\mathrm{sech}\,\xi)}{\omega-\sigma k\,(\tanh\xi\pm i\,\mathrm{sech}\,\xi)}, \tag{4.17}$$

and so on. Starting from $u_6$ obtained in Eq. (3.14) and using BT given in Eq. (4.4) we get

$$u_{45} = \frac{\sigma^2 k^2+\omega^2-2\sigma k\omega\,(\coth\xi\pm\mathrm{csch}\,\xi)}{\omega-\sigma k\,(\coth\xi\pm\mathrm{csch}\,\xi)}, \tag{4.18}$$

and so on. Starting from $u_7$ obtained in Eq. (3.15) and using BT given in Eq. (4.4) we get

$$u_{46} = \frac{\sigma^2 k^2 - \omega^2 - 2\sigma k\omega \,(\tan\xi \pm \sec\xi)}{-\omega - \sigma k\,(\tan\xi \pm \sec\xi)}, \qquad (4.19)$$

and so on. Starting from $u_8$ obtained in Eq. (3.16) and using BT given in Eq. (4.4) we get

$$u_{47} = \frac{-\sigma^2 k^2 + \omega^2 - 2\sigma k\omega \,(\cot\xi \pm \csc\xi)}{\omega - \sigma k\,(\cot\xi \pm \csc\xi)}, \qquad (4.20)$$

and so on. Starting from $u_9$ obtained in Eq. (3.17) and using BT given in Eq. (4.4) we get

$$u_{48} = \frac{(16\sigma^2 k^2 + \omega^2)\tanh^2\xi - 16\sigma k\omega \tanh\xi + 16\sigma^2 k^2 + \omega^2}{\omega \tanh^2\xi - 8\sigma k \tanh\xi + \omega}, \qquad (4.21)$$

Inserting $u_{48}$ into the BT given in Eq. (4.4) we have

$$u_{49} = \frac{(48\sigma^2 k^2\omega + \omega^3)\tanh^2\xi - (24\sigma k\omega^2 + 128\sigma^3 k^3)\tanh\xi + 48\sigma^2 k^2\omega + \omega^3}{(16\sigma^2 k^2 + \omega^2)\tanh^2\xi - 16\sigma k\omega \tanh\xi + 16\sigma^2 k^2 + \omega^2}, \qquad (4.22)$$

and so on. Starting from $u_{10}$ obtained in Eq. (3.18) and using BT given in Eq. (4.4) we get

$$u_{50} = \frac{(16\sigma^2 k^2 - \omega^2)\tan^2\xi + 16\sigma k\omega \tan\xi - 16\sigma^2 k^2 + \omega^2}{\omega - \omega\tan^2\xi + 8\sigma k \tan\xi}, \qquad (4.23)$$

Inserting $u_{50}$ into the BT given in Eq. (4.4) we have

$$u_{51} = \frac{(48\sigma^2 k^2\omega - \omega^3)\tan^2\xi + (24\sigma k\omega^2 - 128\sigma^3 k^3)\tan\xi - 48\sigma^2 k^2\omega + \omega^3}{(16\sigma^2 k^2 - \omega^2)\tan^2\xi + 16\sigma k\omega \tan\xi - 16\sigma^2 k^2 + \omega^2}, \qquad (4.24)$$

and so on. Starting from $u_{11}$ obtained in Eq. (3.19) and using BT given in Eq. (4.4) we get

$$u_{52} = \frac{(16\sigma^2 k^2 - \omega^2)\cot^2\xi - 16\sigma k\omega \cot\xi - 16\sigma^2 k^2 + \omega^2}{\omega - \omega\cot^2\xi - 8\sigma k \cot\xi}, \qquad (4.25)$$

Inserting $u_{52}$ into the BT given in Eq. (4.4) we have

$$u_{53} = \frac{(\omega^3 - 48\sigma^2 k^2\omega)\cot^2\xi + (24\sigma k\omega^2 - 128\sigma^3 k^3)\cot\xi + 48\sigma^2 k^2\omega - \omega^3}{(\omega^2 - 16\sigma^2 k^2)\cot^2\xi + 16\sigma k\omega \cot\xi + 16\sigma^2 k^2 - \omega^2}, \qquad (4.26)$$

and so on. Starting from $u_{12}$ obtained in Eq. (3.20) and using BT given in Eq. (4.4) we get

$$u_{54} = \frac{\sigma^2 k^2 + \omega^2 + (\sigma^2 k^2 + \omega^2)\,\text{sech}\,\xi - 2\sigma k\omega \tanh\xi}{\omega + \omega\,\text{sech}\,\xi - \sigma k \tanh\xi}, \qquad (4.27)$$

and so on. Starting from $u_{13}$ obtained in Eq. (3.21) and using BT given in Eq. (4.4) we get

$$u_{55} = \frac{\sigma^2 k^2 + \omega^2 + (\sigma^2 k^2 + \omega^2)\,i\,\text{csch}\,\xi - 2\sigma k\omega \coth\xi}{\omega + \omega i\,\text{csch}\,\xi - \sigma k \coth\xi}, \qquad (4.28)$$

and so on. Starting from $u_{14}$ obtained in Eq. (3.22) and using BT given in Eq. (4.4) we get

$$u_{56} = -\frac{(\sigma^2 k^2 - \omega^2)\cos\xi + \sigma^2 k^2 - \omega^2 - 2\sigma k\omega \sin\xi}{\omega + \omega\cos\xi + \sigma k \sin\xi}, \qquad (4.29)$$

and so on. Starting from $u_{15}$ obtained in Eq. (3.23) and using BT given in Eq. (4.4) we get

$$u_{57} = \frac{(\sigma^2 k^2 - \omega^2)\sin\xi + \sigma^2 k^2 - \omega^2 + 2\sigma k\omega \cos\xi}{-\omega - \omega\sin\xi + \sigma k \cos\xi}, \qquad (4.30)$$

and so on. Starting from $u_{16}$ obtained in Eq. (3.24) and using BT given in Eq. (4.4) we get

$$u_{58} = \frac{\omega^2 - 4\sigma^2 k^2 - 4\sigma k\omega + (\omega^2 - 4\sigma^2 k^2 + 4\sigma k\omega)\tan\xi}{(\omega + 2\sigma k)\tan\xi + \omega - 2\sigma k}, \tag{4.31}$$

Inserting $u_{58}$ into the BT given in Eq. (4.4) we have

$$u_{59} = \frac{\omega^3 + 8\sigma^3 k^3 - 6\sigma k\omega^2 - 12\sigma^2 k^2\omega + (\omega^3 - 8\sigma^3 k^3 + 6\sigma k\omega^2 - 12\sigma^2 k^2\omega)\tan\xi}{\omega^2 - 4\sigma^2 k^2 - 4\sigma k\omega + (\omega^2 - 4\sigma^2 k^2 + 4\sigma k\omega)\tan\xi}, \tag{4.32}$$

and so on. Starting from $u_{17}$ obtained in Eq. (3.25) and using BT given in Eq. (4.4) we get

$$u_{60} = \frac{\omega^2 - 4\sigma^2 k^2 + 4\sigma k\omega + (\omega^2 - 4\sigma^2 k^2 - 4\sigma k\omega)\cot\xi}{(\omega - 2\sigma k)\cot\xi + \omega + 2\sigma k}, \tag{4.33}$$

Inserting $u_{60}$ into the BT given in Eq. (4.4) we have

$$u_{61} = \frac{\omega^3 - 8\sigma^3 k^3 + 6\sigma k\omega^2 - 12\sigma^2 k^2\omega + (\omega^3 + 8\sigma^3 k^3 - 6\sigma k\omega^2 - 12\sigma^2 k^2\omega)\cot\xi}{\omega^2 - 4\sigma^2 k^2 + 4\sigma k\omega + (\omega^2 - 4\sigma^2 k^2 - 4\sigma k\omega)\cot\xi}, \tag{4.34}$$

and so on. Starting from $u_{18}$ obtained in Eq. (3.26) and using BT given in Eq. (4.4) we get

$$u_{62} = \frac{(\sigma k + \omega)^2 - (\sigma k - \omega)^2 e^\xi}{\sigma k + \omega + (\sigma k - \omega)e^\xi}, \tag{4.35}$$

Inserting $u_{62}$ into the BT given in Eq. (4.4) we have

$$u_{63} = \frac{(\sigma k + \omega)^3 + (\sigma k - \omega)^3 e^\xi}{(\sigma k + \omega)^2 - (\sigma k - \omega)^2 e^\xi}, \tag{4.36}$$

Furthermore, using $u_{63}$ and BT given in Eq. (4.4) we get

$$u_{64} = \frac{(\sigma k + \omega)^4 - (\sigma k - \omega)^4 e^\xi}{(\sigma k + \omega)^3 + (\sigma k - \omega)^3 e^\xi}, \tag{4.37}$$

and so on, we can get a new sequences of exact solution of the space-time CFBE.

Starting from $u_{30}$ obtained in Eq. (3.38) and using BT given in Eq. (4.4) we get

$$u_{65} = \frac{-(\omega^2 + 16\sigma^2 k^2)\cosh\xi\sinh\xi + 8\sigma k\omega\cosh^2\xi - 4\sigma k\omega}{-\omega\cosh\xi\sinh\xi + 4\sigma k\cosh^2\xi - 2\sigma k}, \tag{4.38}$$

and so on, we can get a sequences of solutions generated by the addition of two functions tanh and coth-function solution of the space-time CFBE. Starting from $u_{31}$ obtained in Eq. (3.39) and using BT given in Eq. (4.4) we get

$$u_{66} = -\frac{(\omega^2 - 16\sigma^2 k^2)\cos\xi\sin\xi - 8\sigma k\omega\cos^2\xi + 4\sigma k\omega}{-\omega\cos\xi\sin\xi + 4\sigma k\cos^2\xi - 2\sigma k}, \tag{4.39}$$

and so on, we can get a sequences of solutions generated by the addition of two functions tan and cot-function solution of the space-time CFBE.

By means of the variational iteration method [35] and the Adomian decomposition method [36] the solution of the space-time CFBE in closed form is

$$u_{67} = \frac{(\omega + k) + (\omega - k)\exp(\xi/\sigma)}{1 + \exp(\xi/\sigma)}, \tag{4.40}$$

From $u_{67}$ obtained in Eq. (4.40) we can get a new sequences of exact solution for the space-time CFBE by using BT given in Eq. (4.4) in the form

$$u_{68} = \frac{(\omega+k)^2+(\omega-k)^2 \exp(\xi/\sigma)}{(\omega+k)+(\omega-k) \exp(\xi/\sigma)}, \tag{4.41}$$

$$u_{69} = \frac{(\omega+k)^3+(\omega-k)^3 \exp(\xi/\sigma)}{(\omega+k)^2+(\omega-k)^2 \exp(\xi/\sigma)}, \tag{4.42}$$

$$u_{70} = \frac{(\omega+k)^4+(\omega-k)^4 \exp(\xi/\sigma)}{(\omega+k)^3+(\omega-k)^3 \exp(\xi/\sigma)}, \tag{4.43}$$

$$u_{71} = \frac{(\omega+k)^5+(\omega-k)^5 \exp(\xi/\sigma)}{(\omega+k)^4+(\omega-k)^4 \exp(\xi/\sigma)}, \tag{4.44}$$

and so on, we can get a new sequences of exact solution of the space-time CFBE.

## 5. Conclusion

In this work, we discuss the Painlevé property for non-linear conformal fractional differential equations for the first time . We apply the desired method to the space time conformal fractional Burger's equation. Also, we derive the Bäcklund transform. The general solutions of the space-time conformal CFDEs is described based on the tanh-method, accordingly the method is successively implement to space-time CFBE. Moreover, the space-time CFBE is found to possess the Painlevé property and then Bäcklund transform. Also, we introduced a new recurrence formulae based on Bäcklund transform, that is enable us to derive analytical solution from known solution or old solution to give new solution. New numerous exact solutions are generated based on the Bäcklund transform.

**Acknowledgments**

The authors gratefully acknowledge the approval and the support of this research study by the grant no. SCI-2019-1-10-F-8306 from the Deanship of Scientific Research at Northern Border University, Arar, Saudi Arabia.